 \documentclass[final,5p,times,twocolumn]{elsarticle}
 \usepackage{graphicx}
\usepackage{bm}
\usepackage{txfonts}
\usepackage{hyperref}
\usepackage{mathrsfs}
\usepackage[english]{babel}
\usepackage{amsmath}
\usepackage{cleveref}
\usepackage[autostyle, english = american]{csquotes}
\MakeOuterQuote{"}

\date{\today}
\newcommand{\be}{\begin{equation}}
	\newcommand{\ee}{\end{equation}}

\newcommand{\bfz}{{\bf 0}_{\perp}}

\newcommand{\bfk}{{\bf k}_{\perp}}

\newcommand{\bfkj}{{\bf k}_{\perp j}}

\newcommand{\bfP}{{\bf P}_{\perp}}

\newcommand{\Dp}{{\bf \Delta}_{\perp}}

\usepackage{xcolor}

\usepackage{amssymb}
\usepackage{lipsum}

\journal{Physics Letters B}

\begin{document}

\begin{frontmatter}




\title{Effect of nuclear medium on the spatial distribution of pions}

\author{Satyajit Puhan}
\ead{puhansatyajit@gmail.com}
\author{Navpreet Kaur}
\ead{knavpreet.hep@gmail.com}
\author{Arvind Kumar}
\ead{kumara@nitj.ac.in}
\author{Suneel Dutt}
\ead{dutts@nitj.ac.in}
\author{Harleen Dahiya}
\ead{dahiyah@nitj.ac.in}
\affiliation{Computational High Energy Physics Lab, Department of Physics, Dr. B.R. Ambedkar National Institute of Technology, Jalandhar, Punjab, 144008, India}
\begin{abstract}
We calculate the valence quark generalized parton distributions (GPDs) of the lightest pseudoscalar meson, pion, in an isospin asymmetric nuclear matter at zero temperature by employing a light-cone quark model. The medium modifications in the unpolarized GPDs have been incorporated by taking inputs from the chiral SU($3$) quark mean field model. The electromagnetic form factors (EMFFs) and charge radii have been calculated for both the vacuum and in-medium cases. These results are found to be in agreement with the available experimental data and other model predictions.
\end{abstract}




\end{frontmatter}




\section{Introduction} 
\label{introduction}
Understanding the complex hadron structure through the distribution of their constituents is an interesting topic for the upcoming Electron-Ion collider (EIC) \cite{Accardi:2012qut}. Quarks, gluons and sea quarks, which are supposed to be the internal constituents of a hadron, can be studied using both perturbative and nonperturbative methods. In this perspective, generalized parton distribution functions (GPDs) play an important role in understanding the multi-dimensional structure of a hadron \cite{Diehl:2003ny,Garcon:2002jb,Belitsky:2005qn,Puhan:2024jaw}. The three-dimensional GPDs being a function of longitudinal momentum fraction $(x)$, skewness $(\xi)$ and momentum transferred ($\Delta$) between initial and final state of a hadron, provide the information about charge distributions, form factors, magnetic moments, dipole moments, pressure distributions, shearing force, charge radii, parton distribution functions (PDFs) etc. \cite{Chavez:2021llq,Broniowski:2022iip,Polyakov:2018zvc}. However, GPDs lack in carrying the information of transverse momenta ($\bfk$) of the quark, which can be studied using three-dimensional transverse momentum parton distribution functions (TMDs) \cite{Angeles-Martinez:2015sea,Diehl:2015uka,Puhan:2023ekt}. GPDs can be extracted from deeply virtual Compton scattering (DVCS) \cite{Ji:1996nm} and deeply virtual meson production (DVMP) \cite{Favart:2015umi} processes experimentally. GPDs can be expressed as an off-forward matrix element of light-cone bilocal operators in momentum as well as impact parameter space.  \par
At leading \cite{Dahiya:2007is, Frederico:2009pj, Kumar:2015fta, Kaur:2023zhn} and higher twists \cite{Jain:2024lsj, Sharma:2023ibp, Braun:2023alc}, an enormous amount of work has been done in context of vacuum GPDs. Since nucleon modified structure functions and scattering cross sections of different nuclei have been observed experimentally, one can also expect the medium modified hadron structure when a hadron is surrounded by the nuclear medium. Hence, it is interesting to study the behavior of valence quarks inside a hadron under the effect of asymmetric nuclear medium.
The first indication of variation in the structure function of nucleons in deep inelastic muon scattering on iron and deuterium was observed by European Muon Collaboration (EMC) in 1983 \cite{EuropeanMuon:1983wih}. The Stanford Linear Accelerator Center (SLAC) also provided a benchmark on the effect of baryonic density on deep inelastic scattering (DIS) cross section in different nuclei \cite{Arnold:1983mw}. In context to the modifications induced in the hadronic structure, partial restoration of the chiral symmetry can be  considered as the root cause due to the chiral quark condensation. A concept of partial chiral symmetry restoration is approved experimentally through deeply bound pions \cite{Suzuki:2002ae} and kaons \cite{Sgaramella:2024qdh}. 

\par In this work, we have investigated the effect of isospin asymmetric nuclear medium on the unpolarized valence quark GPDs of pion using light-cone quark model (LCQM) and chiral SU(3) quark mean field (CQMF) model. The relativistic consequences of partonic dynamics within the hadron can be visualized with the help of the light-cone concept, which forms the basis of LCQM. This relativistic framework, which is gauge-invariant, provides a non-perturbative description of the structure and properties of hadrons. LCQM is primarily concerned with valence quarks because they are the primary elements responsible for the overall structure and properties of hadrons. LCQM has accomplished a successful representation of the physical properties of pions in vacuum, such as their electromagnetic properties, charge radii, mass eigenvalues, decay constants, distribution amplitudes, and PDFs \cite{Kaur:2020vkq,Puhan:2023hio}. In CQMF \cite{Wang:2001jw}, the quarks are assumed as the basic building blocks, which are confined inside the baryons via a confining potential and interact through an exchange of the scalar-isovector field $\delta$, the strange scalar-isoscalar field $\zeta$, and the nonstrange scalar-isoscalar field $\sigma$. CQMF has been
successful to calculate the magnetic moments of octet and decuplet baryons in dense nuclear \cite{Singh:2017mxj, Singh:2018kwq} and strange matter \cite{Singh:2020nwp}. 

As the lightest meson, the study of internal structure of pion has always been  fascinating. There are total $8$ GPDs for the case of spin-$0$ mesons up to twist-$4$ \cite{Meissner:2008ay}. At the leading twist, there are total $2$ valence quark GPDs, out of which $F_{1} (x, \xi,-\Delta^2_{\perp})$ is chiral even and  $H_{1} (x, \xi,-\Delta^2_{\perp})$ is chiral odd. In this work, we have limited our calculations to only $F_{1} (x, \xi,-\Delta^2_{\perp})$ GPD as it carries vital information about the electromagnetic form factors (EMFFs), gravitational form factors, charge radius etc. of the pion. This GPD has also been explored in different non-perturbative models like Ads/QCD model \cite{Kaur:2018ewq}, basic light-front quantization (BLFQ) \cite{Adhikari:2021jrh}, non chiral quark mean field model \cite{Son:2024uet}, and Bethe-Salpeter equation (BSE) \cite{Chavez:2021llq}. However, there is no experimental data available for pion GPDs at the present moment, but there have been some lattice simulations results for the unpolarized pion GPDs \cite{Chen:2019lcm,Ding:2024umu}. In our previous calculations on PDFs, TMDs and DAs of pion in asymmetric nuclear medium, we have found out that baryonic density of the medium has a significant effect on valence quark distributions at the model scale \cite{Puhan:2024xdq,Kaur:2024wze}.  Along with this, there have been some work reported on the study of in-medium EMFFs and charge radii for the case of pion in which medium effects are computed through quark-meson coupling (QMC) model and pion structure has been investigated through light-front quark model \cite{Arifi:2024tix}, Nambu–Jona-Lasinio (NJL) model \cite{Hutauruk:2018qku} and BSE \cite{deMelo:2014gea} in symmetric nuclear medium. In-medium gluon and valence quark distributions for pion and kaon have also been studied by using proper-time regularization scheme and NJL model in symmetric nuclear matter \cite{Hutauruk:2021kej}. This is the first time, we are trying to study the behavior of unpolarized pion GPD in the isospin asymmetric nuclear medium. The medium modified EMFFs and charge radii of a pion have also been studied through GPD to investigate the impact of baryonic density of the isospin asymmetric nuclear matter on its spatial structure. 

For the calculations of vacuum quark GPDs, we have used the LCQM \cite{Brodsky:1997de,Acharyya:2024enp,Qian:2008px} in which the quark-quark correlator has been solved to get GPDs in the form of light cone wave functions (LCWFs). In order to incorporate the medium effects of isospin asymmetric nuclear medium, the effective masses, computed through the CQMF model have been used as input parameters in LCQM. 
The $F_{1} (x, \xi,-\Delta^2_{\perp})$ unpolarized GPD, which is denoted as $H(x,\xi,-\Delta^2_\perp)$ for this work, has been solved for $\xi=0$ case, as our main focus is to understand the medium effect on GPDs. The dependence of an unpolarized GPD on the baryonic density and isospin asymmetry of the nuclear medium is presented. Further, vacuum and in-medium EMFFs and charge radii have been computed from an unpolarized GPD of a pion.

This paper is arranged as follows: In Sec.
\ref{SecModel}, we present the details of CQMF and LCQM.
The unpolarized GPD is calculated in Sec. \ref{secgpd}. In Sec. \ref{secres}, the results of present work are discussed, whereas the summary and conclusion are given in Sec. \ref{secsum}.

\section{The models}
\label{SecModel}

\subsection{Chiral SU(3) quark mean field model}\label{cqmf}
This model considers the constituent quarks of a baryon as fundamental degrees of freedom, which are confined within a baryon with confining potential and interact through the scalar ($\sigma$, $\zeta$ and $\delta$) and vector ($\omega$ and $\rho$) fields. Properties of baryons and their constituent quarks are found to be altered due to their interaction with considered fields. The low energy characteristics of QCD such as  spontaneous and explicit breaking of the chiral symmetry are the key ingredients of CQMF model, which have been aroused with the involvement of Lagrangian densities in the form of scalar isoscalar fields $\sigma$ and $\zeta$. The scalar isovector field $\delta$ has also been introduced to study the impact of finite isospin asymmetry. The broken scale invariance property of QCD comes into the picture by taking into account the dilaton field $\chi$.  \par
In order to determine the effective masses of constituent quarks at zero temperature, consider the thermodynamic potential for the isospin asymmetric
nuclear medium \cite{Wang:2001jw}
\begin{eqnarray}
	\Omega &=& - \sum_{i} \frac{\gamma_i}
	{48 \pi^2} \bigg[ (2k^3_{F_i} - 3M^{\ast 2}_i k_{F_i} ) \nu^\ast_i \nonumber \\  &+& 3 m^{\ast 4}_i ln~ \bigg( \frac{k_{F_i} + \nu^\ast_i}{m^\ast_i} \bigg) \bigg] -{\cal L}_{M}-{\cal V}_{\text{vac}} \, , 
	\label{Eq_therm_pot1}  
\end{eqnarray}
where the degeneracy factor is $\gamma_i=2$ (summation $i$ is over the nucleons in the isospin asymmetric nuclear medium) and their Fermi momentum is denoted by $k_{F_i}$.  Vacuum potential energy ${\cal V}_{\text{vac}}$ has been subtracted to attain zero vacuum energy. For given third component of isospin quantum number of nucleons $I^{3i}$, the effective chemical potential of the nucleons in terms of free chemical potential $\nu_i$ is defined as 
\begin{equation}
	\nu_i^\ast = \nu_i - g_{\omega}^i\omega -g_{\rho}^i I^{3i} \rho.
\end{equation}
The quantity $M_i^\ast$ appearing in Eq. (\ref{Eq_therm_pot1}) corresponds to the effective mass of a baryon, which is expressed in terms of its spurious center of momentum and effective energy of the constituent quark $e_q^\ast$ through a relation 
\begin{equation}
	M_i^\ast = 
	\sqrt{\biggl(\sum_q n_{i}^q e_q^\ast + E_{i\,spin} \biggr)^2  - \langle p_{i~cm}^{\ast 2} \rangle} \, ,
\end{equation}
where $n_{i}^q$ represents the number of $q$ flavored quark in the $i^{th}$ baryon \cite{Barik:2013lna}. The term $E_{i~spin}$ is a correction term of baryon energy that attributes the spin-spin interaction and has been adjusted to fit the baryon vacuum mass. In Eq. (\ref{Eq_therm_pot1}), the term ${\cal L}_{M}={\cal L}_{\chi SB}+{\cal L}_{VV}+{\cal L}_{X}$ involves the contribution of the explicit symmetry breaking term ${\cal L}_{\chi SB}$, self-interactions of vector mesons ${\cal L}_{VV}$ and self-interactions of scalar mesons ${\cal L}_{X}$. The explicit expressions of these quantities are
\begin{equation}\label{L_SB}
	{\cal L}_{\chi SB}=\frac{\chi^2}{\chi_0^2}\left[m_\pi^2\kappa_\pi\sigma +
	\left(
	\sqrt{2} \, m_K^2\kappa_K-\frac{m_\pi^2}{\sqrt{2}} \kappa_\pi\right)\zeta\right] \, ,
\end{equation}
\begin{eqnarray}
	{\cal L}_{VV} = \frac{1}{2} \, \frac{\chi^2}{\chi_0^2} \left(
	m_\omega^2\omega^2+m_\rho^2\rho^2\right) + g_4\left(\omega^4+6\omega^2\rho^2+\rho^4\right) \, , 
	\label{vector}
\end{eqnarray}
and
\begin{eqnarray}
	{\cal L}_{X} &=& -\frac{1}{2} \, k_0\chi^2
	\left(\sigma^2+\zeta^2+\delta^2\right)+k_1 \left(\sigma^2+\zeta^2+\delta^2\right)^2
	\nonumber \\ 
	&+&k_2\left(\frac{\sigma^4}{2} +\frac{\delta^4}{2}+3\sigma^2\delta^2+\zeta^4\right) \nonumber \\ &+& k_3\chi\left(\sigma^2-\delta^2\right)\zeta 
	- k_4\chi^4-\frac14\chi^4 {\rm ln}\frac{\chi^4}{\chi_0^4} \nonumber \\ 
	&+&
	\frac{\xi^\prime}
	3\chi^4 {\rm ln}\left(\left(\frac{\left(\sigma^2-\delta^2\right)\zeta}{\sigma_0^2\zeta_0}\right)\left(\frac{\chi^3}{\chi_0^3}\right)\right). \label{scalar0}
\end{eqnarray}
Analogous to the involvement of scalar-isovector field $\delta$ in ${\cal L}_{X}$, the vector-isovector field $\rho$ involved in ${\cal L}_{VV}$, contribute in the medium with finite isospin asymmetry. The logarithmic terms written in Eq. (\ref{scalar0}) represent the scale breaking effects and facilitate the trace anomaly property of QCD in this model. For three colors and flavors, QCD beta-function defines the order of parameter $\xi^\prime$ at one loop level \cite{Papazoglou:1998vr}. Parametrs $k_0, k_1, k_2, k_3$ and $k_4$ have been calculated with the help of $\pi$ meson mass $m_{\pi}$, $K$ meson mass $m_K$ and the average mass of $\eta$ and $\eta'$ meson. The vacuum expectation values of scalar fields $\sigma$ and $\zeta$ are denoted by $\sigma_0$ and $\zeta_0$ respectively. 
The thermodynamic potential for the isospin asymmetric nuclear medium is minimized over the scalar and vector fields and the system of non-linear equations are solved for finite baryonic density and isospin asymmetry of the nuclear medium. The isospin asymmetry in terms of baryonic density $\rho_{B}$ is defined as $\eta =(\rho_n - \rho_p)/2\rho_{B}$, where $\rho_{B}=\rho_n + \rho_p$.

The confinement of quarks within a baryon is defined by the Lagrangian density ${\cal{L}}_c = - \bar{\Psi} \chi_{c} \Psi$ \cite{Singh:2016hiw}, where $\chi_c$ corresponds to the confining potential and can be expressed as 
\begin{equation}\chi_{c}(r) = k_{c}~ r^2(1+\gamma^0)/4.
\end{equation}
Dirac equation for the quark field $\Psi_{qi}$ in presence of confining potential $\chi_{c}(r)$ is defined as follows
\begin{equation}
	\left[-i\vec{\alpha}\cdot\vec{\nabla}+\chi_c(r)+\beta m_q^*\right]
	\Psi_{qi}=e_q^*\Psi_{qi}. \label{Dirac}
\end{equation}
The exchange of scalar and vector fields result into the modification of effective masses and energies of constituent quarks of a baryon, respectively and can be written in terms of their fields as
\begin{equation}
	m_q^*=-g_\sigma^q\sigma - g_\zeta^q\zeta - g_\delta^q I^{3q} \delta + m_0. \label{qmass}
\end{equation}
and 
\begin{equation} 
	e_q^*=e_q-g_\omega^q\omega-g_\rho^q I^{3q}\rho\, ,
	\label{eq_eff_energy1}
\end{equation} 
where $I^{3q}$ is the third component of isospin quantum number of quarks. The Lagrangian density ${\cal L}_{\Delta m} = - (\Delta m)~ \bar \psi S_1 \psi$ introduce the term $m_0$ in Eq. (\ref{qmass}), where  $S_1 \, = \, \frac{1}{3} \, \left(I - \lambda_8\sqrt{3}\right)$ is the matrix for strange $s$ quark. The term $m_0=0$ for $u$ and $d$ quark flavors and takes the value of $77$ MeV for $s$ quark flavor. The coupling terms ($g_\sigma^q$, $g_\zeta^q$, $g_\omega^q$ and $g_\rho^q$) and parameter $k_c$ are adjusted to get the binding energy $-16$ MeV at nuclear saturaton density $0.16$ fm$^{-3}$ \cite{Kumar:2023owb}. 
\subsection{Light-cone quark model}

LCQM is a relativistically crucial framework for explaining the hadronic system in terms of the quarks and gluons degree of freedom that make it up. The expression for the mesonic wave function using the LC Fock-state expansion expressed as \cite{Pasquini:2023aaf}
\begin{eqnarray}
|\mathcal{M}\rangle &=& \sum |q\bar{q}\rangle \psi_{q\bar{q}}
        + \sum
        |q\bar{q}g\rangle \psi_{q\bar{q}g} + \cdots  \, .
\end{eqnarray}
Here $|\mathcal{M}\rangle$ is the meson eigen state.  For this work, we mainly focus on the valence quark distributions by not considering the gluon contribution. So, we continue with $|q\bar{q}\rangle $ state for this work. 

The expansion of meson eigenstate $|\mathcal{M}(P^+,\bfP,S_z)\rangle$ having total momentum $P$ with light-cone coordinates $(P^+,P^-,\bfP)$ and longitudinal spin projection $S_z$ can be expressed in terms of multiparticle Fock eigenstates $|n\rangle$ as \cite{Qian:2008px,Lepage:1980fj}
\begin{eqnarray}
|\mathcal{M}(P^+,\bfP,S_z)\rangle= \sum_{n,\lambda_j} \int \prod_{j=1}^{n} \frac{dx_j~  d^2\bfkj}{2(2\pi)^3\sqrt{x_{j}}} \, 16 \pi^{3} \, \nonumber \\
 \delta ~ \bigg(1-\sum_{j=1}^{n} x_{j}\bigg) \, \delta^{(2)} \bigg(\sum_{j=1}^{n}\bfkj\bigg) \nonumber \\		
 \psi_{n/\mathcal{M}}(x_{j},\bfkj,\lambda_{j})|n; x_{j} P^{+},x_{j}\bfP + \bfkj,\lambda_{j}\rangle \, ,
\label{MesonState}\end{eqnarray}
where $x_j=\frac{k_j^+}{P^+}$ with $0 \leq x_j \leq 1$ is the longitudinal momentum fraction of the $j$th constituent parton with $\bfkj$ and $\lambda_j$ as its  transverse momentum and helicity respectively. For pseudo-scalar mesons $S_z=0$.
The multiparticle state of $n$-particles is normalized as 
\begin{eqnarray}
\langle n; k^{\prime +}_j, \bfkj^\prime, \lambda_{j}^\prime|n ; k^+_j, \bfkj, \lambda_j \rangle = \prod_{j=1}^{n} 16 \pi^{3} \,  k^{\prime +}_j \, \delta (k^{\prime +}_j-k^+_j) \, \nonumber \\
 \delta^{(2)} ( \bfkj^\prime-\bfkj) \, \delta_{\lambda_{j}^\prime \lambda_{j}} \, .
\end{eqnarray}
The two particle Fock-state for the case of pion can be expressed as 
\begin{eqnarray}
|\mathcal{\pi} (P^+,\bfP,S_z=0) =\sum_{\lambda_1,\lambda_2}\int
\frac{\mathrm{d} x \mathrm{d}^2
        \mathbf{k}_{\perp}}{\sqrt{x(1-x)}16\pi^3} \, \nonumber \\
           \psi(x,\mathbf{k}_{\perp},\lambda_1, 
          \lambda_2)|x, x  
 \mathbf{P}_\perp+\mathbf{k}_{\perp},
        \lambda_1,\lambda_2 \rangle
        .
        \label{meson}
\end{eqnarray}
Here $\lambda_{1(2)}$ are the helicities of quark (antiquark) respectively. $\psi(x,\mathbf{k}_{\perp},\lambda_1, 
          \lambda_2)$ is the total wave function of pion. With all possible helicities of quark and antiquark, the pion state can be expressed as 
\begin{eqnarray}
|\mathcal{\pi} (P^+,\bfP,S_z=0)\rangle &=& \int \frac{dx \, d^2 \bfk}{  16 \pi^3 \sqrt{x(1-x)}} \, \nonumber \\ &\times& \big[ \psi (x,\bfk,\uparrow,\uparrow) \, |x P^+, x  
 \mathbf{P}_\perp+\mathbf{k}_{\perp}, \uparrow, \uparrow \rangle   \nonumber \\
&+& \psi (x,\bfk,\uparrow,\downarrow) \, |x P^+, x  
 \mathbf{P}_\perp+\mathbf{k}_{\perp}, \uparrow, \downarrow \rangle  \nonumber \\
&+& \psi (x,\bfk,\downarrow,\uparrow) \, |x P^+, x  
 \mathbf{P}_\perp+\mathbf{k}_{\perp},  \downarrow,\uparrow \rangle \nonumber \\ &+& \psi (x,\bfk,\downarrow,\downarrow) \, |x P^+,x  
 \mathbf{P}_\perp+\mathbf{k}_{\perp}, \downarrow, \downarrow \rangle \big] \, . \nonumber \\ \label{eqnq} 
\end{eqnarray}
The momenta of meson and its constituent quarks $u$($\bar{d}$), having effective  masses $M_{\pi}^*$ and $m_u^*$($m_{\bar{d}}^*$) respectively, in the light-cone frame are expressed as
\begin{eqnarray}
P &=& \bigg(P^+,\frac{M_{\pi}^{\ast2}}{P^+},\bfz\bigg) \, ,  \nonumber \\
k_1 &=& \bigg(x P^+,\frac{\bfk^2 + m^{\ast2}_u}{x P^+},\bfk \bigg) \, ,  \nonumber \\
k_2 &=& \bigg((1-x) P^+,\frac{\bfk^2 + m^{\ast2}_{\bar{d}}}{(1-x) P^+},-\bfk \bigg) \, . 
\end{eqnarray}
The bound state meson mass $M_{\pi}^*$ of pion is expressed as 
\be
M_{\pi}^* = \sqrt{\frac{m_u^{*2}+\bfk^2}{x}+\frac{m_{\bar d}^{*2}+\bfk^2}{(1-x)}} \, .
\ee
Here the active quark carries $x$ fraction of longitudinal momentum and $(1-x)$ momentum fraction carried by the antiquark to obey the momentum conservation rule. The total wave function $\psi(x,\bfk,\lambda_1,\lambda_2)$ ion Eq. (\ref{meson}) of pion can be expressed as the product of spin and momentum space wave function as \cite{Huang:1994dy} 
\be 
\psi(x,\bfk,\lambda_1, \lambda_2)= \varphi(x,\bfk) \, \Phi (x,\bfk,\lambda_1, \lambda_2) \, .\label{sm}
\ee 
Here $\varphi(x,\bfk)$ is the momentum space wave function and has been  expressed using Brodsky-Huang-Lepage prescription as 
\cite{Kaur:2020vkq,Xiao:2002iv,Yu:2007hp} 
\begin{eqnarray} 
\varphi (x,\bfk) &=& \mathcal{A} \, exp \,\Biggl[-\frac{ \frac{m_{u}^{\ast2} + \bfk^2}{x} + \frac{m_{\bar{d}}^{\ast2} + \bfk^2}{1-x}}{8 \beta^2_k} \nonumber \\ &-& \frac{(m_u^{\ast2} - m_{\bar{d}}^{\ast2})^2}{8 \beta^2_k \, \bigg( \frac{m_u^{\ast2} + \bfk^2}{x} + \frac{m_{\bar{d}}^{\ast2} + \bfk^2}{1-x}\bigg)}\Biggr] \, ,
\end{eqnarray}
where $\mathcal{A}= A \, exp \, \big[\frac{m_u^{\ast2} + m_{\bar{d}}^{\ast2}}{8 \beta^2_k}\big]$ with $A$ and $\beta_k$ representing the normalization constant and harmonic scale parameter respectively. The momentum space wave function is normalized as
\begin{equation}
    \int \frac{{d x} d^2 \bfk}{2 (2 \pi)^3} \, |\varphi (x,\bfk)|^2 =1 \, .
\end{equation}

The spin wave function $\Phi(x,\bfk,\lambda_1, \lambda_2)$ in Eq. (\ref{sm}) is obtained by solving the quark meson vertex as \cite{Choi:1996mq,Qian:2008px}
\begin{equation}
    \Phi(x,\textbf{k}_\perp, \lambda_1, \lambda_2) =  \frac{\bar u (k_1,\lambda_1)~\gamma_5~ v(k_2,\lambda_2)}{\sqrt{2}\sqrt{M_\pi^{*2}-(m^\ast_{u}-m^\ast_{\bar d})^2}} \, .
\end{equation}
Here $u$ and $v$ are the Dirac spinors. While solving the above equation, we found the different spin wave function for different polarization as 
\begin{equation}
  \begin{array}{lll}
    \Phi(x,\mathbf{k}_\perp,\uparrow,\uparrow)&=&\frac{1}{\sqrt{2}}\omega^{-1}(-\textbf{k}^L)(M^{*2}_\pi+m^*_u+m^*_{\bar d}),\\
    \Phi(x,\mathbf{k}_\perp,\uparrow,\downarrow)&=&\frac{1}{\sqrt{2}}\omega^{-1}((1-x)m^*_u+x m^*_{\bar d})(M^{*2}_\pi+m^*_u+m^*_{\bar d}),\\
    \Phi(x,\mathbf{k}_\perp,\downarrow,\uparrow)&=&\frac{1}{\sqrt{2}}\omega^{-1}(-(1-x)m^*_u-x m^*_{\bar d})(M^{*2}_\pi+m^*_u+m^*_{\bar d}),\\
    \Phi(x,\mathbf{k}_\perp,\downarrow,\downarrow)&=&\frac{1}{\sqrt{2}}\omega^{-1}(-\textbf{k}^{R})(M^{*2}_\pi+m^*_u+m^*_{\bar d}) \, .
  \end{array}.
\end{equation}
Here $\omega=(M^*_\pi+m^*_u+m^*_{\bar d})\sqrt{x(1-x)[M_{\pi}^{*2}-(m^*_u-m^*_{\bar d})^2]}$ and $k^{L(R)}= k^{1}\pm \iota k^{2}$. The above spin wave function can be obtained by transforming instant-form SU($6$) wave function into light-cone form by making the use of Melosh-Wigner rotation and will have the same form.

\par The in-medium longitudinal momentum fraction ($x^*$) of $j$th quark/antiquark is related to vacuum longitudinal momentum fraction ($x$) by the relations \cite{Puhan:2024xdq}
 \begin{align}
 x_j^*  = 
 \begin{cases}
 \frac{E_j^* + g_{\omega}^{j}\omega + 
 g_{\rho}^{j} I^{3j}\rho + k_j^{*3}}{E_j^* + E_{\bar j}^* + g_{\rho}^{{ j}} \left(I^{3{j}}-I^{3{\bar j}}\right)\rho + P^{*3}} = \frac{x_j+ (g_{\omega}^{j}\omega + 
 g_{\rho}^{j} I^{3j}\rho)/P^+}{1+\left(I^{3{j}}-I^{3{\bar j}}\right)\rho/P^+} \quad \text{for quark } q \\
  \frac{E_{\bar j}^* - g_{\omega}^{j}\omega - 
  g_{\rho}^{j} I^{3\bar j}\rho + k_{\bar j}^{*3}}{E_j^* + E_{\bar j}^* + g_{\rho}^{{ j}} \left(I^{3{j}}-I^{3{\bar j}}\right)\rho + P^{*3}} = \frac{x_j - (g_{\omega}^{j}\omega + 
 g_{\rho}^{j} I^{3j}\rho)/P^+}{1+\left(I^{3{j}}-I^{3{\bar j}}\right)\rho/P^+} \quad
  \text{for antiquark } \bar{q} .
 \end{cases}
 \label{Eq_xfrac_med2}
 \end{align}
However, for the sake of simplicity, we have constrained our results to $x$, instead of $x^*$.

\section{Generalized parton distributions (GPDs)}\label{secgpd}
GPDs are evaluated through the matrix elements of quark operators at a light-like separation \cite{Diehl:2003ny}. For pseudoscalar mesons, we have only one chiral-even unpolarized GPD at leading twist which can be defined in terms of the bilocal current as
\begin{eqnarray}
    H(x, \xi, t) &=& \frac{1}{2} \int \frac{dz^-}{2\pi} e^{ix\bar{P}^+z^-} \bigg\langle \mathcal{M}(P^\prime,\lambda^\prime)~\bigg|~ \bar{\vartheta} (0)~ \nonumber \\
    &\times& W(0,z) \gamma^+ ~ \vartheta (z) ~\bigg| \mathcal{M}(P,\lambda) \bigg\rangle \, . 
    \label{gpd}
\end{eqnarray}
Here, position four vector is denoted by $z=(z^+,z^-,z_\perp)$. $\vartheta (z^-)$ and $W(0,z)$ are the quark field operator and the Wilson line (which connects the two quark field operators), respectively. However, in this work, Wilson line is taken as unity. $P$ and $P^\prime$ are the initial and final state meson momentum. Other kinematic variables which include the four-momentum transfer and skewness parameter are respectively expressed as $\Delta^\mu=P^{\prime \mu}-P^\mu$ with $t=\Delta^2=-\Delta^2_\perp$ and $\xi=-\Delta^+/2P^+$. For this work, we have taken $\xi=0$. By substituting pion state wave function of Eq. (\ref{MesonState}) in the above equation, the valence quark GPD $H(x,0,t)$ is found to be 

\begin{align}
    \notag&H\left(x,0, t\right)  =\int \frac{d^{2} \bfk}{16 \pi^{3}} \sum_{\lambda_{q_{i}}} \sum_{\lambda_{q_{f}}} \\\notag&\times\left[\psi^{*}\left(x, \bfk^{\prime\prime},\lambda_{q_{f}},\uparrow\right) \psi\left(x, \bfk^{\prime},\lambda_{q_{i}},\uparrow\right)
    \right.\\\notag&\left.+\psi^{*}\left(x, \bfk^{\prime\prime},\lambda_{q_{f}},\downarrow\right) \psi\left(x, \bfk^{\prime},\lambda_{q_{i}},\downarrow\right) \right]
    \\&\times\frac{u_{\lambda_{q_{f}}}^{\dagger}\left(x P^{+}, \bfk-\frac{\boldsymbol{\Delta}_{\perp}}{2}\right) \gamma^{0} \gamma^{+} u_{\lambda_{q_{i}}}\left(x P^{+}, \bfk+\frac{\boldsymbol{\Delta}_{\perp}}{2}\right)}{2 x P^{+}}\, .
\end{align}
Here $\lambda_{f}$ and $\lambda_{i}$ are the final and initial helicity of the active quark of pion respectively. The above equation in the overlap form of LCWFs is found to be
\begin{eqnarray}
    H(x,0,t)=\int \frac{d^2 \mathbf{k_\perp}}{16 \pi^3} \big[\psi^\ast (x^{\prime\prime},\mathbf{k}^{\prime\prime}_\perp, \uparrow, \uparrow)  \psi (x^{\prime},\mathbf{k}^{\prime}_\perp, \uparrow, \uparrow) 
    \nonumber \\ 
     + \psi^\ast (x^{\prime\prime},\mathbf{k}^{\prime\prime}_\perp, \uparrow, \downarrow) \psi (x^{\prime},\mathbf{k}^{\prime}_\perp, \uparrow, \downarrow) \nonumber \\ 
   + \psi^\ast (x^{\prime\prime},\mathbf{k}^{\prime\prime}_\perp, \downarrow, \uparrow) \psi (x^{\prime},\mathbf{k}^{\prime}_\perp, \downarrow, \uparrow) \nonumber \\ + \psi^\ast (x^{\prime\prime},\mathbf{k}^{\prime\prime}_\perp, \downarrow, \downarrow) \psi (x^{\prime},\mathbf{k}^{\prime}_\perp, \downarrow, \downarrow)\big] \, ,
\label{GPDeq}
\end{eqnarray}
where final and initial state quark momentum are represented by $\bfk^{\prime\prime}$ and $\bfk^{\prime}$, respectively. In symmetric frame, they have a form
\begin{eqnarray}
    \bfk^{\prime\prime}&=&\bfk-(1-x^{\prime\prime})~\frac{\Dp}{2} \, , \nonumber \\
    \bfk^{\prime}&=&\bfk+(1-x^{\prime})~\frac{\Dp}{2} \, .
\end{eqnarray}
For zero skewness, the initial and final state longitudinal momentum fraction carried by an active quark of a meson remain the same i.e. $x^{\prime\prime}=x^{\prime}$. Hence, we have written initial and final state longitudinal momentum fraction by $x$ only.
The contribution of $q$-quark flavor to the total elastic EMFF of meson can be evaluated by the zeroth moment of the unpolarized GPD $H(x,0,t)$, which can be expressed as
\be
F^u(Q^2)=\int dx ~ H(x,0,t) \, .
\ee
where $Q^2=-t$. The total EMFF of pion can be obtained by taking account of both the quark and antiquark contributions as 
\begin{eqnarray}
F(Q^2)&=& F^u(Q^2)+F^{\bar d}(Q^2) \nonumber\\
&=&
e_{u} \int dx ~H(x,0,t) + e_{\bar d} \int dx ~H(1-x,0,t) \, , \nonumber \\
\end{eqnarray}
where $e_{u(\bar d)}$ are the charge of $u$ quark and $\bar d$ antiquark. The pion EMFFs obey the sum rule of $F(Q^2=0)=1$ at every baryonic density.
\par The corresponding mean square charge radius of pion can be computed using
\begin{equation}
\langle r^2_\pi \rangle= -6 \frac{\partial F(Q^2)}{\partial Q^2}\Big|_{Q^2\rightarrow0} \, .
\end{equation}
The charge radii of pion with different quark flavor can be written as 
\begin{equation}
  \langle r^2_\pi \rangle=e_u \langle r^2_u \rangle + e_{\bar d} \langle r^2_{\bar d} \rangle \, .
\end{equation}
With this expression, we can find the radii of different flavor. However for the case of pion, both flavor have same radii.
\par Similarly the unpolarized valence quark PDF of pion can be obtained at $t=0$ as 
\be
f(x)= H(x,0,0) \, .
\ee

\section{Results and discussion}\label{secres}
\begin{figure*}
	\centering
	\begin{minipage}[c]{0.98\textwidth}
		(a)\includegraphics[width=7.5cm]{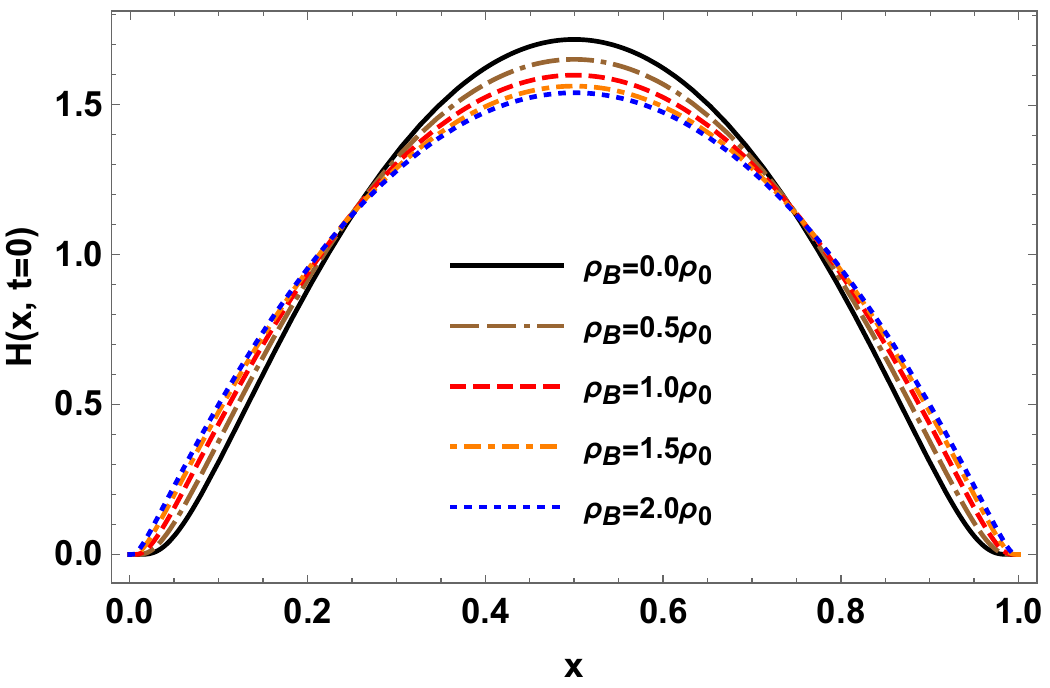}
		\hspace{0.07cm}	
		(b)\includegraphics[width=7.5cm]{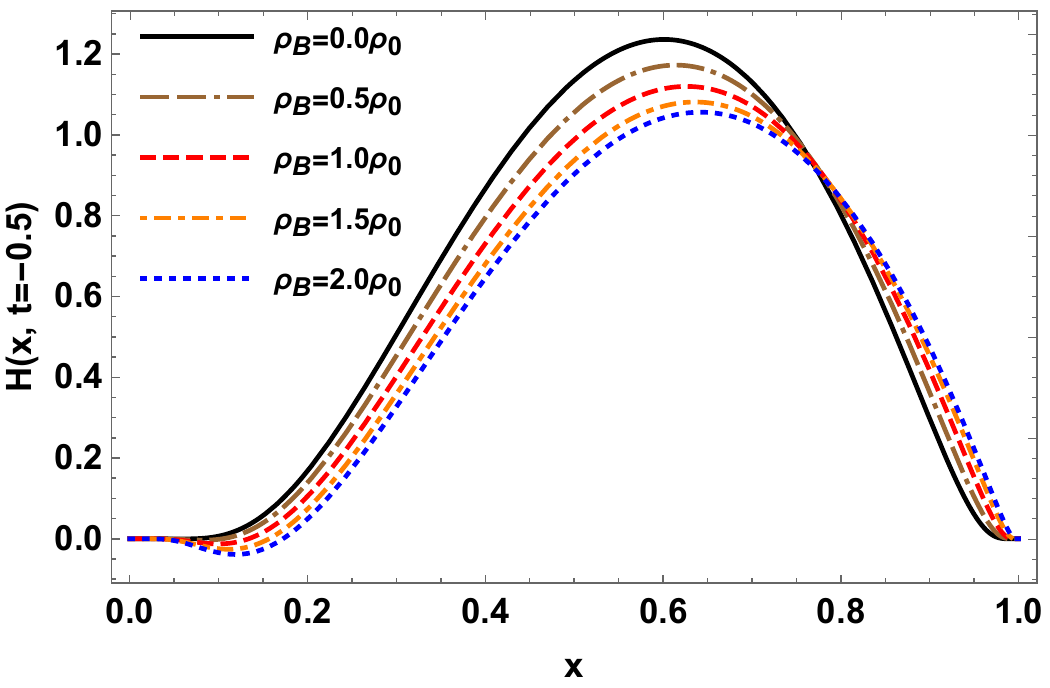}
		\hspace{0.07cm}\\
		(c)\includegraphics[width=7.5cm]{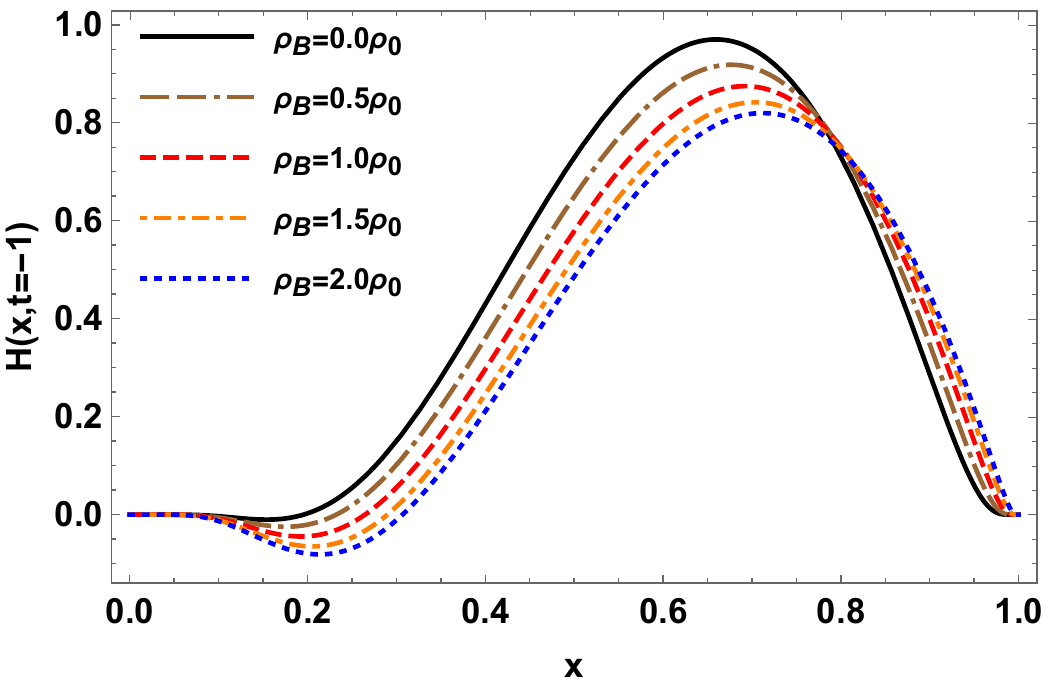}
		\hspace{0.03cm} 
		(d)\includegraphics[width=7.5cm]{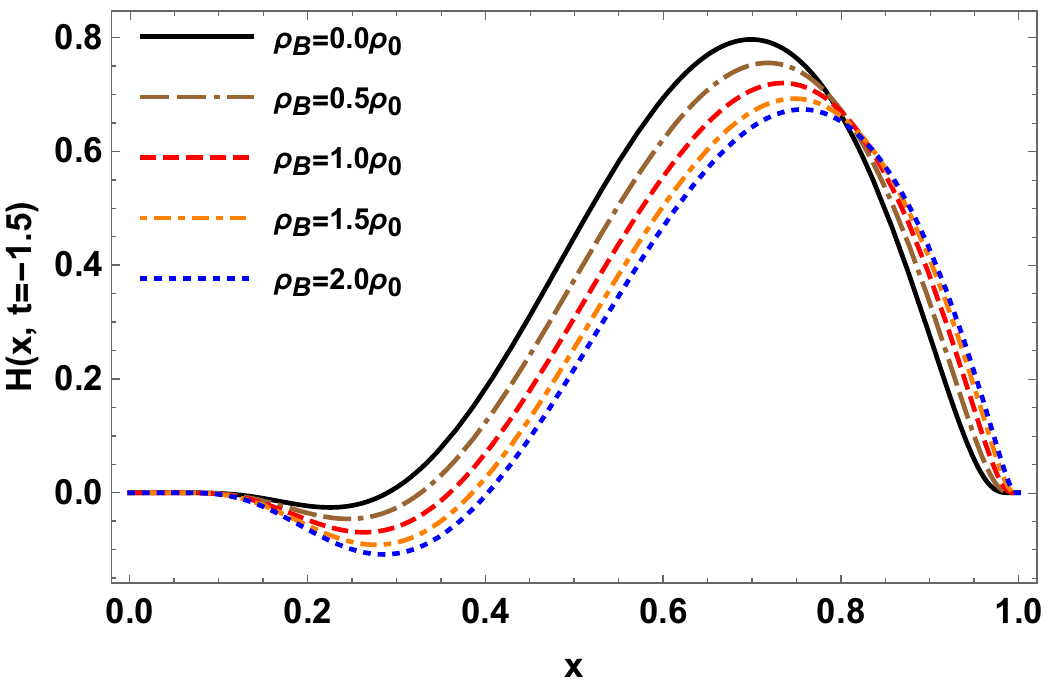}
		\hspace{0.03cm}	
	\end{minipage}
	\caption{\label{GPDx} (Color online) Vacuum and in-medium unpolarized pion GPD $H(x, t)$ with baryonic density upto $\rho_B=2 \rho_0$ as a function of longitudinal momentum fraction $x$ for four discrete values of  transverse momentum transfer (a) $t=0$  GeV$^2$, (b) $t=-0.5$  GeV$^2$, (c) $t=-1.0$  GeV$^2$ and (d) $t=-1.5$  GeV$^2$.}
\end{figure*}

For numerical predictions, we require the effective constituent quark masses $m_{u(\bar{d})}$ of a pion and a harmonic scale parameter $\beta_k$ as input parameters. For the case of vacuum in LCQM, these parameters have been fitted with the mass and decay constant of pion and have been taken from Ref. \cite{Puhan:2024xdq}. The results of unpolarized GPD of pion have been portrayed in Fig. \ref{GPDx} as a function of longitudinal momentum $x$ for different finite values of invariant transverse momentum transfer at $-t$ (GeV$^2$) $=0, 0.5, 1$ and $1.5$ in subplots ($a-d$) respectively. Vacuum distribution of a pion shows a symmetric behavior at $t=0$. However, as the value of $-t$ increases from $0$ to $1.5$ with an interval of $0.5$, the amplitude of distribution reduces with narrowing down over smaller region of $x$ and the peak is shifted towards higher values of $x$. This implies that if the transfer of transverse momentum increases, an active quark has more chance to carry higher longitudinal momentum fraction. Similar kind of dependency has also been observed in results obtained from AdS/QCD inspired holographic model \cite{Kaur:2018ewq} and BSE \cite{Albino:2022gzs}. In order to analyze the impact of baryonic density $\rho_{B}$ on the distributions, we have taken its value upto $2 \rho_{0}$ (with step size of $0.5$) and plotted the juxtapose of vacuum and in-medium distributions for each finite value of $-t$ in Fig. \ref{GPDx}. At $t=0$, the distributions of pion GPDs (generally, termed as PDFs) become flatter and broader with an increase of $\rho_{B}$ as presented in Fig. \ref{GPDx} (a). More specifically, the amplitude of an in-medium distribution is found to be more for the regions $0<x<0.25$ and $0.75<x<1$, whereas it is found to be less in between these regions with increase of baryonic density. Fig. \ref{GPDx} (b) portrays the distribution at $-t=0.5$, and a comparison of vacuum with in-medium distributions at finite values of $\rho_{B}$ up to $2 \rho_{0}$ implies that the magnitude of in-medium distributions reduce with drifting of peak to larger values of $x$ faster till $1.5 \rho_{0}$. However, beyond $x=0.77$, the amplitude of in-medium distribution enhances and is also found to be negative for the smaller region at low $x$. On further increase in the momentum transfer during the process, similar kind of trend has been followed by the GPDs with more pronounced negative distribution on smaller region of longitudinal momentum fraction $x$ as the baryonic density increases. \par
\begin{figure*}
	\centering
	\begin{minipage}[c]{0.98\textwidth}
		(a)\includegraphics[width=7.5cm]{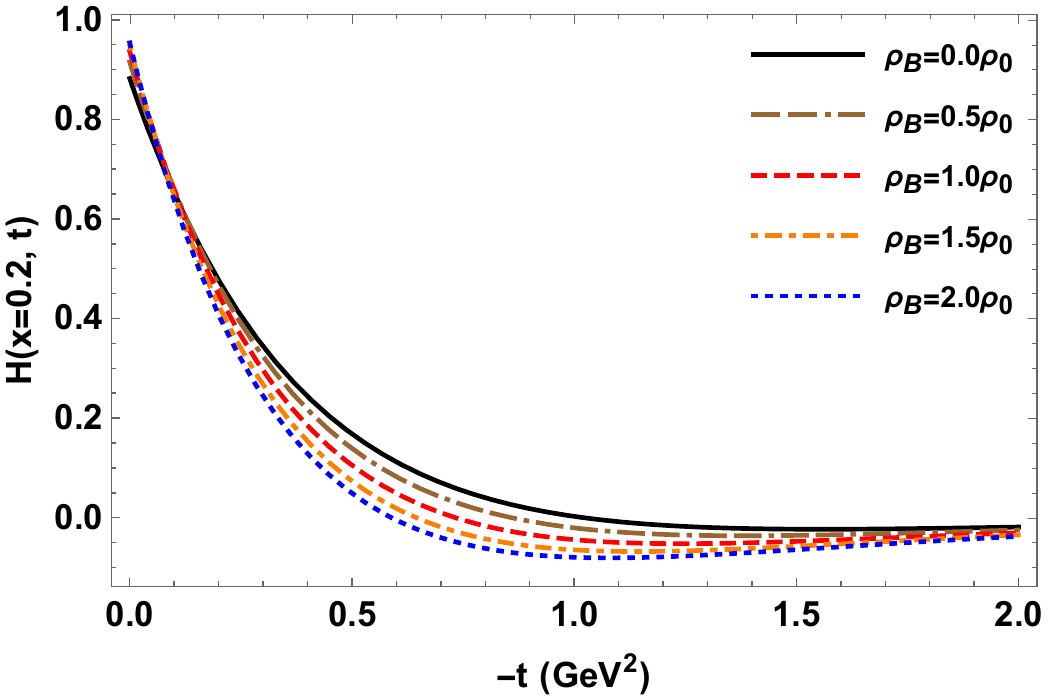}
		\hspace{0.03cm}	
		(b)\includegraphics[width=7.5cm]{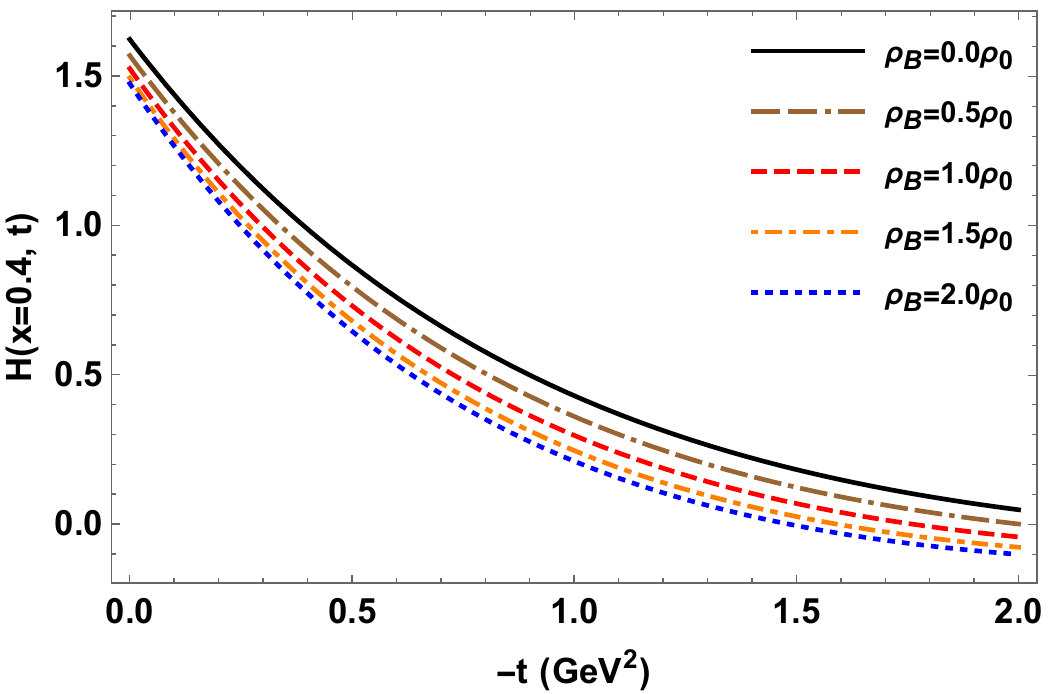}
		\hspace{0.03cm}\\
		(c)\includegraphics[width=7.5cm]{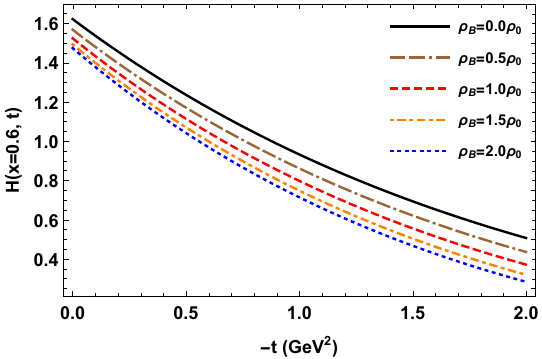}
		\hspace{0.03cm} 
		(d)\includegraphics[width=7.5cm]{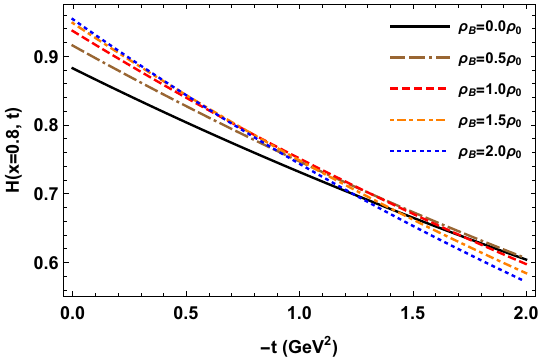}
		\hspace{0.03cm}	
	\end{minipage}
	\caption{\label{GPDt} (Color online) Vacuum and in-medium unpolarized pion GPD $H(x, t)$ with baryonic density upto $\rho_B=2 \rho_0$ as a function of transverse momentum transfer $-t$ (GeV$^2$) for four discrete values of longitudinal momentum fraction (a) $x=0.2$, (b) $x=0.4$, (c) $x=0.6$ and (d) $x=0.8$.}
\end{figure*}

\begin{figure*}
	\centering
	(a)\includegraphics[width=7.5cm]{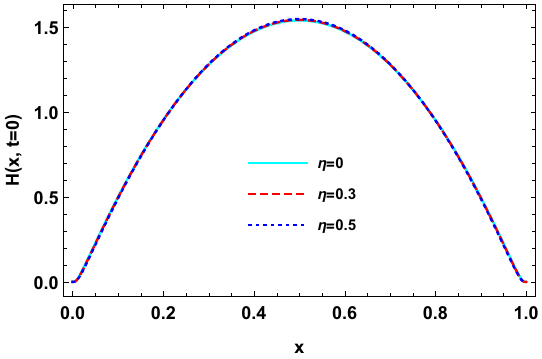}
	\hspace{0.03cm}			(b)\includegraphics[width=7.5cm]{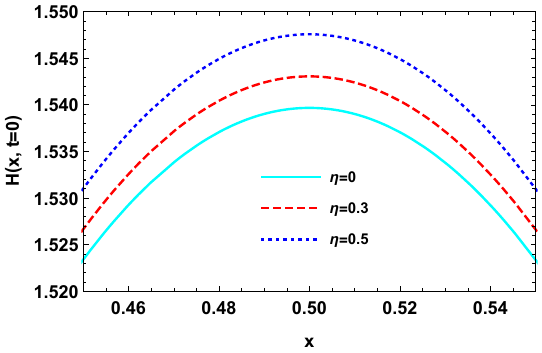}
	\hspace{0.03cm}\\
	(c)\includegraphics[width=7.5cm]{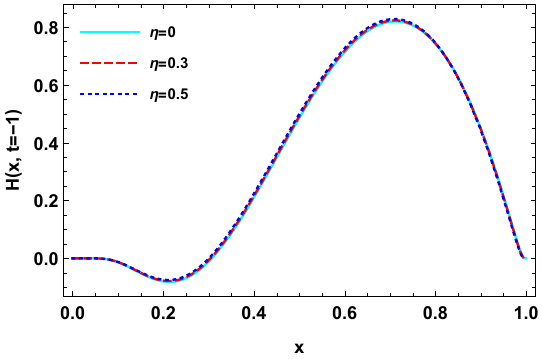}
	\hspace{0.03cm} 
	(d)\includegraphics[width=7.5cm]{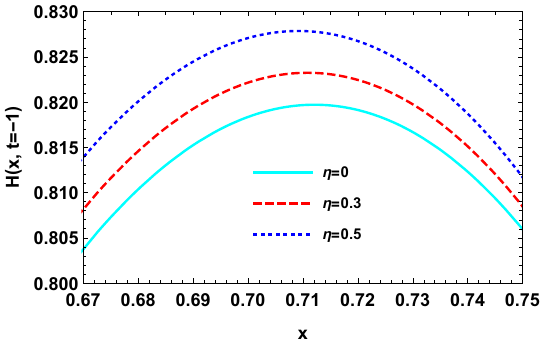}
	\hspace{0.03cm} 
	\caption{\label{GPDAsym} (Color online) In-medium unpolarized pion GPD $H(x, t)$ as a function of longitudinal momentum fraction $x$ for different values of isospin asymmetry parameter ($\eta=0, 0.3$ and $0.5$) and fixed value of baryonic density $\rho_B=2 \rho_0$ at transverse momentum transfer (a) $t=0$  GeV$^2$ and (c) $t=-1$  GeV$^2$ with their zoomed plots around the region of peak values in subplots (b) and (d), respectively.}
\end{figure*}

 
The dependence of unpolarized GPDs of pion on $-t$ for fixed values of longitudinal momentum fraction $x$ has been represented in Fig. \ref{GPDt} ($a-d$) for $x=0.2, 0.4, 0.6$ and $0.8$ respectively. As the value of $x$ increases discretely, the distribution becomes steeper. The juxtapose of vacuum and in-medium distributions represent that for $x=0.2, 0.4$ and $0.6$, the in-medium distributions fall off more rapidly than vacuum distributions as a function of $-t$ and the rapidness increases as the medium becomes more denser (such that $\rho_{B}$ increases). This is in line of what is expected because the rapid fall off corresponds to more interactions with the nuclear medium. However, at $x=0.8$, we found that higher the baryonic density, more steeper is the slope of a distribution as a function of $-t$. Effect of isospin asymmetry parameter $\eta$ has been represented in Fig. \ref{GPDAsym} (a) on GPD at zero transverse momentum transfer i.e. $t$ (GeV$^2$) $=0$ and fixed value of baryonic density $\rho_B=2 \rho_0$. In our previous works \cite{Puhan:2024xdq, Singh:2024lra}, we found that the $\eta$ parameter shows significant difference only at higher baryonic densities, that's why we fixed baryonic density $\rho_B=2 \rho_0$ to analyze its effect on GPDs and observed a similar over all trend. Its zoomed plot has been represented in Fig. \ref{GPDAsym} (b) around its peak value and we found that only the magnitude of GPDs slightly increases with increase in the value of $\eta$ which is due to the slower rate of declination in the effective quark masses when the isospin asymmetry enhances in the nuclear medium. Similar pattern can be observed for higher value of $-t$ which has been represented in subplots (c) and (d) of Fig. \ref{GPDAsym}.  \par
\begin{figure*}
\centering
\begin{minipage}[c]{0.98\textwidth}
(a)\includegraphics[width=5.45cm]{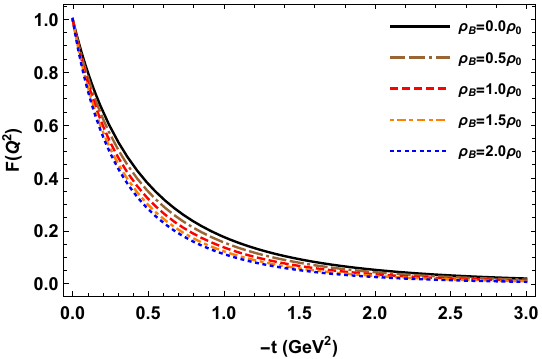}
\hspace{0.03cm}	
(b)\includegraphics[width=5.45cm]{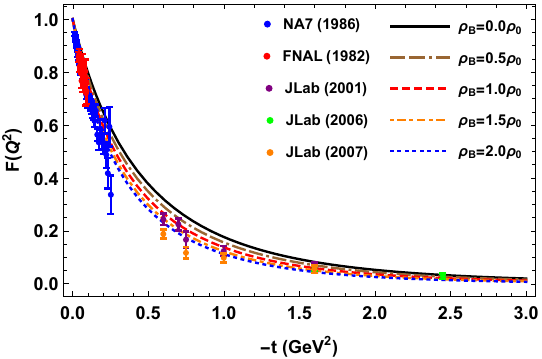}
\hspace{0.03cm}
(c)\includegraphics[width=5.45cm]{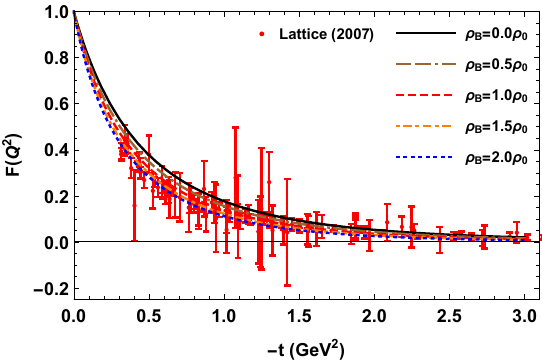}
\hspace{0.03cm}
\end{minipage}
\caption{\label{FF} (Color online) Comparison of electromagnetic form factors among vacuum and in-medium distributions with baryonic density upto $\rho_B=2 \rho_0$ in (a), along with experimental available data in (b) and lattice simulated data in (c).}
\end{figure*}

Fig. \ref{FF} ($a$) represents the results of vacuum and in-medium EMFFs, which shows that with increase in the value of $\rho_{B}$, the EMFFs drop down more rapidly as a function of $-t$ (GeV$^2$). It implies that the interactions become more pronounced when the nuclear medium becomes more and more denser, resulting into the larger spatial spread of an electric charge that will also be seen in the results of an electric charge radii. Their comparison with available experimental and lattice simulation data has been demonstrated in Figs. \ref{FF} ($b$) and ($c$), respectively and the results are found to be in good agreement. 
From the comparison of vacuum and in-medium charge radii, presented in Table \ref{table-moment}, we observe that the absolute values of the in-medium charge radius increase
with increasing nuclear density. Their comparative analysis has also been presented in the same table with in-medium predictions from other model results \cite{Gifari:2024ssz,Arifi:2024tix} and vacuum results from available experimental \cite{Amendolia:1984nz} and lattice simulation \cite{Gao:2021xsm} data. The qualitative trend of an increment in the value of $\sqrt{\langle r^2_\pi \rangle}$) is found to be same with the results of Refs. \cite{Gifari:2024ssz,Arifi:2024tix}. These outcomes also align with previous studies that employed in different theoretical approaches \cite{deMelo:2014gea,Yabusaki:2023zin}. However, it should be kept in mind that there exists a recent theoretical prediction
showing the pion charge radius might start to decrease
at very high nuclear densities, as indicated in Ref. \cite{Yabusaki:2023zin}, although additional experimental confirmation might be needed to validate this tendency. When the quark mass decreases kinematically, the quarks can travel across a wider spatial region and the quark antiquark binding energy decreases, meaning the pion is less bound, increasing the charge radius.
\section{Summary and conclusions}\label{secsum}
In the present paper, we have studied the effect of asymmetric nuclear medium on the valence quark distributions of the lightest pseudoscalar meson (pion) by evaluating in-medium generalized parton distributions (GPDs). The in-medium effects have been introduced into GPDs by treating effective quark masses as input parameters. These input parameters have been calculated by using chiral SU(3) quark  mean field model. Through this model, we found that as the medium around pion meson becomes denser, the effective masses of quark (antiquark) reduces. Further, light-cone dynamics has been used to investigate the valence quark distributions in light-cone quark model. Comparison between vacuum and in-medium GPDs as a function of longitudinal momentum fraction $x$ for finite invariant transverse momentum transfer $-t$ suggest that with increase of baryonic density $\rho_{B}$, the distributions get concentrated on the smaller region of $x$, but at comparatively higher values of $x$. However, at zero transverse momentum transfer, the distribution flattens down with expansion over large region of $x$ with increase of $\rho_{B}$. An effect of isospin asymmetry parameter $\eta$ represents an increase in the amplitude of the distribution with an enhancement of isospin asymmetry in the nuclear medium. The distribution of GPDs as a function of $-t$ for fixed values of $x$ fall off more rapidly for denser nuclear medium than vacuum due to the reduction of effective quark masses in nuclear medium. The results of vacuum electromagnetic form factor (EMFF) of pion are found to be in agreement with the available experimental data and the in-medium EMFFs are found to fall off more rapidly than vacuum, suggesting larger spatial spread of valence quark charge distribution, which is consistent with the results aroused for vacuum and in-medium electric charge radii. The observed vacuum and in-medium spatial information of pion will provide fruitful insights for the future measurements of pion in J-PARC in Japan \cite{Aoki:2021cqa} and EIC in US \cite{AbdulKhalek:2021gbh}.

\begin{table}
\centering
\begin{tabular}{|c|c|c|c|}
\hline 
Baryonic density & \multicolumn{1}{c|}{LCQM } & \multicolumn{1}{c|}{NJL Model}& \multicolumn{1}{c|}{LFQM}\\
 ratio $(\rho_B/\rho_0)$ & \multicolumn{1}{c|}{(This work)} & \multicolumn{1}{c|}{\cite{Gifari:2024ssz}}& \multicolumn{1}{c|}{\cite{Arifi:2024tix}} \\
\cline{2-4}
 & $\sqrt{\langle r^2_\pi \rangle}$ (fm) & $\sqrt{\langle r^2_\pi \rangle}$ (fm) & $\sqrt{\langle r^2_\pi \rangle}$ (fm) \\
\hline
$0$ &  0.523 & 0.629 & 0.654\\
$0.5$ &  0.553 & 0.664 & 0.776 \\
$1$ &  0.585 & 0.694 & 0.947\\
$1.5$ &  0.614 & 0.714 & -  \\
$2$ &  0.638 & 0.730 & - \\
Exp. \cite{Amendolia:1984nz} &  0.653(10) & - & -  \\
Lat. \cite{Gao:2021xsm} &  0.648(2) & - & - \\

\hline
\end{tabular}
\caption{The observed charge radii of pion at different baryonic density in LCQM and their comparison with other model predictions \cite{Gifari:2024ssz,Arifi:2024tix} along with experimental data \cite{Amendolia:1984nz} and lattice simulation data \cite{Gao:2021xsm}.}
\label{table-moment}
\end{table}



\section*{Acknowledgements}
H.D. would like to thank the Science and Engineering Research Board, Anusandhan-National Research Foundation, Government of India under the scheme SERB-POWER Fellowship (Ref. No. SPF/2023/000116) for financial support.
\appendix



\end{document}